\documentclass[prd,amsmath,amssymb,superscriptaddress,floatfix,nofootinbib,10pt]{revtex4}
\usepackage{times}
\usepackage{amssymb,amsbsy,amsmath,amsfonts}
\usepackage{graphicx}
\usepackage{float}
\usepackage{color}
\usepackage{morefloats}
\usepackage{rotating}
\usepackage{srcltx}
\usepackage{slashed}
\usepackage{multirow}
\usepackage{verbatim}
\usepackage{hyperref}
\usepackage{tabularx}
\usepackage{bm}
\allowdisplaybreaks[4]

\usepackage{bbding}
\usepackage{threeparttable}

\newcommand{\PreserveBackslash}[1]{\let\temp=\\#1\let\\=\temp}
\newcolumntype{C}[1]{>{\PreserveBackslash\centering}p{#1}}
\newcolumntype{R}[1]{>{\PreserveBackslash\raggedleft}p{#1}}
\newcolumntype{L}[1]{>{\PreserveBackslash\raggedright}p{#1}}

\begin{document}

\title{$B^+$ decay to $K^+\eta\eta$ with ($\eta\eta$) from the $D\bar{D}(3720)$ bound state}

\begin{abstract}
We search for a $B$ decay mode where one can find a peak for a $D \Bar{D}$ bound state predicted in effective theories and in Lattice QCD calculations, which has also been claimed from some reactions that show an accumulated strength in $D \Bar{D}$ production at threshold. We find a good candidate in the $B^+\to K^+ \eta\eta$ reaction, by looking at the $\eta\eta$ mass distribution. The reaction proceeds via a first step in which one has the $B^+\to D_s^{*+} \Bar{D}^0$ reaction followed by $D_s^{*+}$ decay to $D^0 K^+$ and a posterior fusion of $D^0 \Bar{D}^0$ to $\eta \eta$, implemented trough a triangle diagram that allows the $D^0 \Bar{D}^0$ to be virtual and produce the bound state. The choice of $\eta\eta$ to see the peak is based on results of calculations that find the $\eta\eta$ among the light pseudoscalar channels with stronger coupling to the $D \Bar{D}$ bound state. We find a neat peak around the predicted mass of that state in the $\eta\eta$ mass distribution, with an integrated branching ratio for $B^+\to K^+$ ($D\Bar{D}$, bound) ; ($D\Bar{D}$, bound) $\to \eta \eta$ of the order of $1.5 \times 10^{-4}$, a large number for hadronic $B$ decays, which should motivate its experimental search.
\end{abstract}


\date{\today}
\author{Pedro C. S. Brandão}
\email[E-mail me at: ]{pedro.brandao@ufba.br}
\affiliation{Instituto de Física, Universidade Federal da Bahia, Campus Ondina, Salvador, Bahia 40170-115, Brazil}
\affiliation{Departamento de Física Teórica and IFIC, Centro Mixto Universidad de Valencia-CSIC Institutos de Investigación de Paterna, 46071 Valencia, Spain}

\author{Jing Song}
\email[E-mail me at: ]{Song-Jing@buaa.edu.cn}
\affiliation{School of Physics, Beihang University, Beijing, 102206, China}
\affiliation{Departamento de Física Teórica and IFIC, Centro Mixto Universidad de Valencia-CSIC Institutos de Investigación de Paterna, 46071 Valencia, Spain}

\author{Luciano M. Abreu}
\email[E-mail me at: ]{luciano.abreu@ufba.br}
\affiliation{Instituto de Física, Universidade Federal da Bahia, Campus Ondina, Salvador, Bahia 40170-115, Brazil}
\affiliation{Instituto de F\'{\i}sica, Universidade de S\~{a}o Paulo, 
Rua do Mat\~{a}o, São Paulo SP, 05508-090, Brazil}

\author{ E.Oset}
\email[E-mail me at: ]{oset@ific.uv.es}
\affiliation{Departamento de Física Teórica and IFIC, Centro Mixto Universidad de Valencia-CSIC Institutos de Investigación de Paterna, 46071 Valencia, Spain}
\maketitle

\maketitle
\section{\label{sec:level1}Introduction}

The search for hadronic states in the charm sector and the description of their structure is attracting much attention recently as evidenced by the large amount of review papers devoted to the subject~\cite{Esposito:2014rxa,Lebed:2016hpi,Chen:2016qju,Guo:2017jvc,Kalashnikova:2018vkv,Yamaguchi:2019vea,Brambilla:2019esw,Guo:2019twa,Chen:2022asf,Mai:2022eur}. We mention two examples: the state $X(3872)$ couples strongly to $D^{*}\Bar{D}$ and it is a subject of debate concerning its nature as a $D^{*}\Bar{D}$ molecule or a compact tetraquark state~\cite{Song:2023pdq}, and the $T_{cc}(3875)$~\cite{LHCb:2021vvq,LHCb:2021auc}, coupling strongly to $DD^{*}$, is also thought to be a $DD^{*}$ molecule state, but others opinions have also being given (see list of references in~\cite{Dai:2023kwv}). Taking advantage of this wave of enthusiasm on this subject we want to come back to a recurrent problem, the possible existence of a $D\Bar{D}$ bound state, proposing a method to find it experimentally. The state was predicted studying the meson-meson interaction in the charm sector in~{\cite{Gamermann:2006nm}} and was bound about $20$ MeV. The state was confirmed in posterior theoretical studies~\cite{Nieves:2012tt,Hidalgo-Duque:2012rqv}. More recently it was also found in lattice calculations~\cite{Prelovsek:2020eiw}. \\

Several works have tried to see experimental evidence for its existence. Since $D\Bar{D}$ bound cannot decay into meson states containing $c\Bar{c}$, evidence for its existence has been searched for in the $D\Bar{D}$ distribution  close to the threshold in several reactions. In~\cite{Gamermann:2007mu} support for its existence was found in the $ee^{-}\rightarrow D\Bar{D}$ reaction looking at the $D\Bar{D}$ spectrum close to the threshold. An updated experimental work for this reaction was done in~\cite{Belle:2017egg} and, again, support for the $D\Bar{D}$ state from this reaction and $\gamma\gamma\rightarrow D\Bar{D}$ was claimed in~\cite{Wang:2019evy,Wang:2020elp}. A more refined theoretical work of these two latter reactions was done in~\cite{Deineka:2021aeu} claiming evidence for this bound state. In~\cite{Xiao:2012iq} three reactions were proposed to observe this bound state, but none has being done so far to found it. In~\cite{Dai:2020yfu,Gamermann:2009ouq} it was suggested to be found in the $\psi(3770)$ radiative decay,  $\psi(3770) \to \gamma D^{0}\Bar{D}^{0}$, and in~\cite{Dai:2015bcc} in the $B^{+}\to D^{0}\Bar{D}^0K^+$ and $B^{0}\to D^{0}\Bar{D}^0K^{0}$ decays, looking in both cases for the $D^{0}\Bar{D}^0$ mass distribution close to threshold.\\

In the present work we propose a different reaction, the $B^{+}\rightarrow K^{+}\eta\eta$, looking for the $\eta\eta$ invariant mass distribution where a peak is being expected. The reason to propose this reaction is double. Among the light pairs of light pseudoscalar mesons into which this state could decay, the $\eta\eta$  channel stands as one of the most important. On the other hand, in the PDG~\cite{ParticleDataGroup:2022pth} one finds that  the reaction $B^{+}\rightarrow D_{s}^{*+}\Bar{D}^0$ has  a very large branching fraction for a $B^{+}$ decay, of the order of $10^{-2}$. It might seem that this decay has nothing to do with the $K^{+}\eta\eta$ decay, but we will show that a triangle diagram with $B^{+}\rightarrow D_{s}^{*+}\Bar{D}^0$ followed by $D_{s}^{*+}\rightarrow D^{0}K^{+}$, and fusion of $\Bar{D}^{0}D^0$ to produce the $D\Bar{D}$ bound final state, with its posterior decay to $\eta\eta$, has a reasonable large branching fraction which would make this decay easily accessible. 

The $D\Bar{D}$ bound state with isospin $I=0$ looks now more acceptable after the discovery of the $T_{cc}(3875)$, but from the theoretical point of view, it resembles very much the $f_{0}(980)$, which couples strongly to $K\Bar{K}$, that has been obtained within the chiral unitary approach~\cite{Oller:1997ti,Kaiser:1998fi,Locher:1997gr,Pelaez:2006nj,Hanhart:2008mx,Nieves:1998hp,Nieves:2009ez}. Since a general rule is that the binding of states becomes larger when going from lighter to heavier quarks with the same configuration~\cite{Ader:1981db}, the existence of the $D\Bar{D}$ bound state seems unavoidable, and with this conviction we propose the new reaction with the $B^{+}\rightarrow K^{+}\eta\eta$ decay which is accessible by the LHCb and Belle collaborations.

\section{formalism}
The idea is to find an efficient mechanism to produce $\eta\eta$ at the end. For this purpose it is not necessary to produce $\eta\eta$ in a first step in a $B$ decay. Instead, the idea is to produce $D\Bar{D}$ since this is the main component of the $D\Bar{D}$ bound state and $\eta\eta$ is only one decay channel. Yet, we wish to have three particles in the final state (including $\eta\eta$) because then we can play with the $\eta\eta $ invariant mass and observe the peak of the $D\Bar{D}$ bound state. The idea is then to produce one particle and $D\Bar{D}$. Then the $D\Bar{D}$ can interact producing the $D\Bar{D}$ bound state. One way to accomplish it is to produce $D_s^{*+}\Bar{D}^0$, let  $D_s^{*+}$ decay to $K^+D^0$ and then we have the pair $D^0\Bar{D}^0$ to interact and proceed via $D\Bar{D}\to \eta\eta$.\\

\begin{figure}[H]
    \centering
\includegraphics[width=0.44\textwidth]{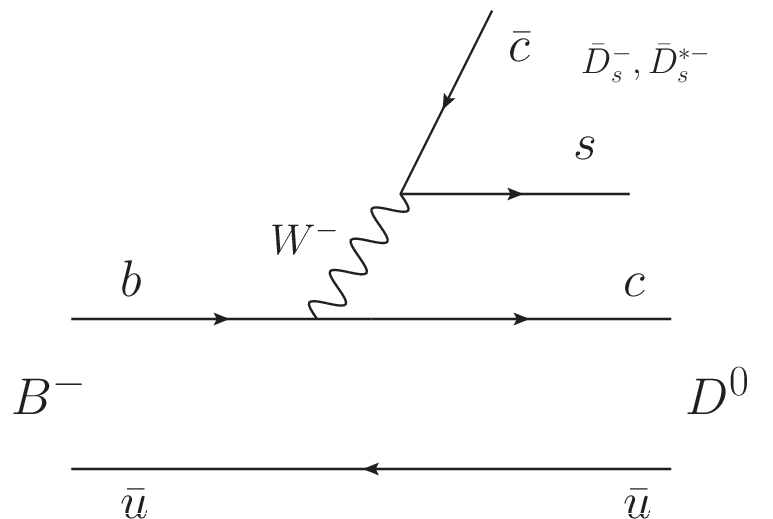}
    \caption{Cabibbo favored process for a $B$ decay, with external emission at the quark level.}
    \label{exter}
\end{figure}

The choice of the first step is most welcome since the process proceeds via the most Cabibbo favored mode for a $B$ decay, with external emission, as shown in Fig.~\ref{exter} for the complex conjugate $B^-\to D_s^{*-}D^0$ reaction. This favors a large rate of this decay mode and one finds the branching fraction~\cite{ParticleDataGroup:2022pth},

\begin{align}\label{eq1}
Br[B^+\to D_{s}^{*+}\Bar{D}^0]=(7.6\pm1.6)\times 10^{-3}.
\end{align}

\begin{figure}[H]
    \centering
\includegraphics[width=0.44\textwidth]{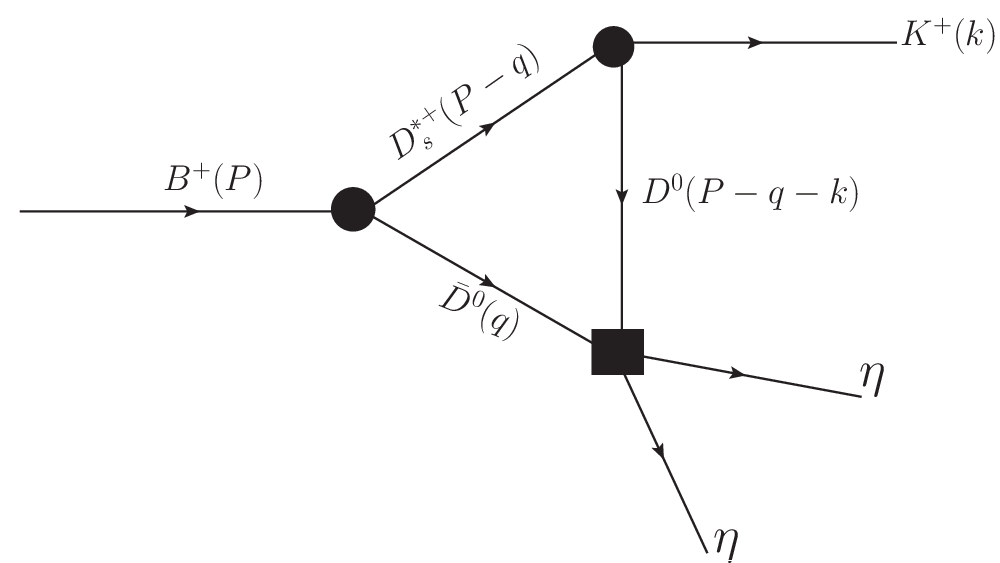}
\caption{Mechanism producing $\eta\eta$ through the rescattering of ${D}^0\Bar{D}^0$.}
    \label{mechanismq}
\end{figure}

This is a big rate for a $B$ decay,  which necessarily involves a suppressed Cabibbo transition $b \to c$.
The next step after the {$D_s^{*+}\Bar{D}^0$} production is to allow the $D_s^{*+}$ decay to $D^0K^+$(virtually) and then proceed with the $D^0\Bar{D}^0$ transition to $\eta\eta$, where the peak of the bound state would show up. This process is depicted in Fig.~\ref{mechanismq}, through a triangle diagram, which, however, does not develop a triangle singularity~\cite{Landau:1959fi}, since $D_s^{*+}\to D^0K^+$ is kinematically forbidden and one cannot place the three intermediate particles on shell~\cite{Bayar:2016ftu}.

We take the meson masses from the PDG~\cite{ParticleDataGroup:2022pth},
\begin{align}
m_{{B}^+} = 5279.34~\mathrm{MeV},& \quad
m_{\eta} = 547.862~\mathrm{MeV}, \quad
m_{{D}^0} = 1864.84~\mathrm{MeV}, \\\nonumber
m_{{K}^+}  = &~ 493.677~\mathrm{MeV}, \quad
\mathrm{and}~ M_{ D_{s}^{*+}} = 2112.2~\mathrm{MeV}.
\end{align}

\subsection{$B^+$ decay to   $D_{s}^{*+}\Bar{D}^0$}
In the diagram of Fig.~\ref{mechanismq} we have a vertex  $D_{s}^{*+}\to K^+{D}^0$ which one can obtain from a standard Lagrangian, the ${D}^0\Bar{D}^0\to \eta\eta$ scattering amplitude that one takes from~\cite{Xiao:2012iq} and the  $B^+ \to D_{s}^{*+}\Bar{D}^0$ transition, determined from the experiment as described   below.\\

The $B^+\to D_{s}^{*+}\Bar{D}^0$ vertex has the typical structure of a vector coupling to two pseudoscalars as follows
\begin{align}
    t_1 = C \epsilon^\mu(P+q)_\mu, 
\end{align}
where $\epsilon^\mu$ is the polarization vector of the $D_{s}^{*+}$, and $C$ the coupling constant.\\

The $B^+ \to D_{s}^{*+}\Bar{D}^0$ width is given by
\begin{align}
    \Gamma[B^+\to D_{s}^{*+}\Bar{D}^0]
    =& \frac{1}{8\pi} \frac{1}{m^2_B} \sum_{pol} |t_1|^2 q,
\end{align}
with
\begin{align}
q=\frac{\lambda^{1/2}(m^2_B,m^2_{\Bar{D}^0},m_{D^*_s}^2)}{2m_B}.
\end{align}

After some algebra, we obtain
\begin{align}
    \sum_{pol} |t_1|^2  =& C^2 \sum_{pol}  \epsilon^\mu(P+q)_\mu   \epsilon^\nu(P+q)_\nu       =  4 C^2 \Bigg(\frac{m_B}{m_{D^*_s}}\Bigg)^2 \Vec{q}~^2,
\end{align}
then the branching fraction can be written as
\begin{align}
    Br[B^+\to D_{s}^{*+}\Bar{D}^0]
    =\frac{\Gamma}{\Gamma_B} =  \frac{1}{\Gamma_B}\frac{1}{2\pi} C^2\frac{q^3}{m_{D^*_s}^2} ,
\end{align}
and using Eq.~(\ref{eq1}) we find
\begin{align}\label{Csover}
    \frac{C^2}{\Gamma_B}
    = \frac{2\pi m_{D^*_s}^2}{q^3}~(7.6\pm1.6)\times 10^{-3}=0.00537~\text{MeV}^{-1}.
\end{align}

\subsection{loop  evaluation  }
We now evaluate the amplitude for the $K^+\eta\eta$ triangle diagram of Fig.~\ref{mechanismq}. We  construct the $D_s^{*+}\to {D}^0K^{+}$ vertex  by using the effective Lagrangian:
\begin{align}\label{trace}
    \mathcal{L}_{VPP}= -ig \langle [P,\partial_\mu P] V^\mu\rangle,
\end{align}
where $g=\frac{m_\mathrm{v}}{2f_\pi}$, $m_\mathrm{v}=800$~MeV, $f_\pi=93$~MeV, and $P,~V$ are the $q\Bar{q}$ matrices with $u,~d,~s$ quarks, written in terms of pseudoscalar ($P$) or vector mesons ($V$) as 

\begin{align}\label{meson_P}
\centering
P=
\left(
  \begin{array}{cccc}
    \frac{\pi^0}{\sqrt{2}}+\frac{\eta}{\sqrt{3}}+\frac{\eta'}{\sqrt{6}} & \pi^{+} & K^{+} & \Bar{D}^{0}\\
    \pi^{-} & \frac{-\pi^0}{\sqrt{2}}+\frac{\eta}{\sqrt{3}}+\frac{\eta'}{\sqrt{6}} & K^0 & D^{-}\\
    K^{-} & \bar{K}^0 & -\frac{\eta}{\sqrt{3}}+\sqrt{\frac{2}{3}}\eta' & D_{s}^{-} \\
    D^{0} & D^{+} & D_{s}^{+} & \eta_{c} \\
  \end{array}
\right),~~~~
\end{align}
\begin{align}\label{meson_V}
\centering
V_{\mu}=
\left(
  \begin{array}{cccc}
    \frac{\rho^0}{\sqrt{2}}+\frac{\omega}{\sqrt{2}} & \rho^{+} & K^{*+} & \Bar{D}^{*0} \\
    \rho^{-} & \frac{-\rho^0}{\sqrt{2}}+\frac{\omega}{\sqrt{2}} & K^{*0} & D^{*-} \\
    K^{*-} & \bar{K}^{*0} & \phi & D_{s}^{*-}\\
    {D}^{*0} & D^{*+} & D_{s}^{*+} & J/\Psi \\
  \end{array}
\right)_{\mu},
\end{align}
where we have taken the ordinary $\eta-\eta'$ mixing of Ref.~\cite{Bramon:1992kr}. The symbol $\langle...\rangle$ in Eq.~(\ref{trace}) denotes the trace in $SU(4)$. Note  {however}, that with this {algorithm} one is only making use of the $q\Bar{q}$ character of the meson~\cite{Sakai:2017avl}.

We find for this vertex 
\begin{align}
    -it_2 =  -ig \epsilon^\mu[ (2k-P+q)_\mu] 
\end{align}
Then, the loop amplitude is given by

\begin{align}
    -it_L =& \int \frac{d^4q}{(2\pi)^4}(-i) C \epsilon^\mu[ (P+q)_\mu] g(-i)\epsilon^\nu[ (2k-P+q)_\nu] \times (-i)t_{{D}^0\Bar{D}^0,\eta\eta}(M_\mathrm{{inv}}(\eta\eta))  \\ \nonumber 
    & \times\frac{i}{q^2-m_{D^0}^2+i\epsilon} \times\frac{i}{(P-q)^2-m_{D^*_s}^2+i\epsilon} \times\frac{i}{(P-q-k)^2-m_{D^0}^2+i\epsilon},  \\ \nonumber          
\end{align}
where $M_\mathrm{{inv}}(\eta\eta)$ is the invariant mass of the $\eta\eta$ system. 
By doing the sum over polarizations for the vector meson $D_s^{*}$ we get 
\begin{align}
    \sum \epsilon^\mu[ (P+q)_\mu] \epsilon^\nu[ (2k-P+q)_\nu]
    =& \Bigg[-g^{\mu\nu}+\frac{(P-q)^\mu(P-q)^\nu}{m_{D^*_s}^2}\Bigg](P+q)_\mu(2k-P+q)_\nu\\ \nonumber
    =& m_B^2-q^2-2Pk-2kq+ \frac{1}{m_{D^*_s}^2}\Bigg[(m_B^2-q^2)(2Pk-2kq-m_B^2-q^2+2Pq)\Bigg].     
\end{align}

We perform the $q^0$ integration analytically using Cauchy's residues. For this purpose we use 
\begin{align}
    \frac{i}{q^2-m_{D^0}^2+i\epsilon} = \frac{1}{2w_D(q)}\Bigg(\frac{1}{q^0-w_{D}(q)+i\epsilon}-\frac{1}{q^0+w_{D}(q)-i\epsilon}\Bigg),
\end{align}
and  keep only the positive energy part because we are dealing with heavy particles. The Cauchy integration picks up the pole $q^0=w_{D^0}(q)$.
As a consequence, we obtain the loop amplitude 
\begin{align}\label{t_loop}
    t_L =& \int \frac{d^3q}{(2\pi)^3} C g \Bigg[-m_{D^0}^2-m_K^2+M^2_\mathrm{{inv}}(\eta\eta)-2w_k(k)w_D(q)+2\Vec{k}\cdot\Vec{q} \\ \nonumber  
     & + \frac{1}{m_{D^*_s}^2}(m_B^2-m_{D^0}^2)\Bigg(m_K^2-m_{D^0}^2-M^2_\mathrm{{inv}}(\eta\eta)-2w_k(k)w_D(q)+2\Vec{k}\cdot\Vec{q}+2m_Bw_D(q)\Bigg)\Bigg]  \\ \nonumber 
    & \times t_{{D}^0\Bar{D}^0,\eta\eta}(M_\mathrm{{inv}}(\eta\eta)) \times \frac{1}{2w_D(q)}  \times \frac{1}{2w_{D^*_s}(q)}   \times \frac{1}{2w_D(\Vec{k}+\Vec{q})}  \\ \nonumber    
    &\times  \frac{1}{m_B-w_D(q)-w_{D^*_s}(q)+i\epsilon} \times \frac{1}{m_B-w_D(q)-w_k(k)-w_{D}(\Vec{k}+\Vec{q})+i\epsilon}, \quad F_\mathrm{HQS}\Theta(q_\mathrm{max}-|\Vec{q}~^*|),
    \end{align}
with
\begin{align}
k=\frac{\lambda^{1/2}(m^2_B,m^2_k,M^2_\mathrm{{inv}}(\eta\eta))}{2m_B},
\end{align}
where we have used
\begin{align}
    &(P-k)^2= M^2_\mathrm{{inv}}(\eta\eta), \\ \nonumber 
    &2Pk= m_B^2+m_k^2-M^2_\mathrm{{inv}}(\eta\eta), \\ \nonumber 
    &2kq= 2w_k(k)w_D(q)-2\Vec{k}\cdot\Vec{q}, \\ \nonumber 
    &2Pq= 2m_Bw_D(q).
\end{align}

In Eq.~(\ref{t_loop}) we have the factor $F_\mathrm{HQS}$ given by 
\begin{align*}
    F_\mathrm{HQS}=\frac{m_{D^*_s}}{m_{k^*}},
\end{align*}
which stems from consideration of heavy quark spin symmetry~\cite{Liang:2014eba} to obtain correct width of the   $D^*\to D\pi$ decay. On the other hand the factor $\Theta(q_\mathrm{max}-|\Vec{q}~^*|)$ comes from the way that we regularize the loops in our $t$ amplitudes with the cut-off method, which implies that the $t$ matrix has the structure~\cite{Gamermann:2009uq}
\begin{align}
    t(\Bar{q},\Bar{q~}')=t\Theta(q_\mathrm{max}-|\Vec{q}|)\Theta(q_\mathrm{max}-|\Vec{q}~'|),
\end{align}
where $q_\mathrm{max}$ is the regulator in the loop functions $G$ in the $T=[1-VG]^{-1}V$ matrix. Since $q_\mathrm{max}$ regulates the $D\Bar{D}$ loops in their rest frame, we must take the factor $\Theta(q_\mathrm{max}-|\Vec{q}~^*|)$ where $\Vec{q}~^*$ is the $\Bar{D}^0$ momentum in the rest frame of $\eta\eta$ given by~\cite{Bayar:2016ftu} as

\begin{align}
    \Vec{q}~^*=\Bigg[\Bigg(\frac{E_R}{M_\mathrm{{inv}}(\eta\eta)}-1\Bigg)\frac{\Vec{q}\cdot\Vec{k}}{\Vec{k}^2}+\frac{w_{D^0}(q)}{M_\mathrm{{inv}}(\eta\eta)}\Bigg]\Vec{k}+\Vec{q},
\end{align}
with 
\begin{align}
E_R=\sqrt{M^2_\mathrm{{inv}}(\eta\eta)+\Vec{k}^2}.
\end{align}
Now the integral only depends on $|\Vec{k}|$, and hence on $M_\mathrm{{inv}}(\eta\eta)$. \\

The matrix elements for the $D\Bar{D}\to j$ transition are obtained using the Bethe-Salpeter equation $T=[1-VG]^{-1}V$ in coupled channels in Ref.~\cite{Xiao:2012iq}, but since the couplings of the state to the $D\Bar{D}$ bound state are calculated there, we directly take the $t_{{D}^0\Bar{D}^0,\eta\eta}$ transition amplitude from this reference and write it with a  Breit-Winger form as
\begin{align}\label{BW}
t_{{D}^0\Bar{D}^0,\eta\eta}(M_\mathrm{{inv}}(\eta\eta))= \frac{g_{{D}^0\Bar{D}^0}g_{\eta\eta} }{M^2_\mathrm{{inv}}(\eta\eta)-m^2_{{D}^0\Bar{D}^0}+im_{{D}^0\Bar{D}^0}\Gamma_{{D}^0\Bar{D}^0}},
\end{align}
with the relevant quantities given by~\cite{Xiao:2012iq}
\begin{align}
&g_{{D}^0\Bar{D}^0} = (5962 + i~1695)~\mathrm{MeV}, \\ \nonumber
&g_{\eta\eta} =(1023 +i~24)~\mathrm{MeV}, \\ \nonumber
 &m_{{D}\Bar{D}}|_b = 3722~\mathrm{MeV}~(\mathrm{for~the~bound~state}), \\ \nonumber
&\Gamma_{{D}\Bar{D}}|_b =36~\mathrm{MeV}. \\ \nonumber
\end{align}

Finally, the differential mass distribution for the $\eta\eta$ system is given by  
\begin{align}
\frac{d\Gamma}{dM_\mathrm{{inv}}(\eta\eta)}=\frac{1}{(2\pi)^3}\frac{1}{4m_B^2}k\Tilde{P_\eta}|t_L|^2,
\end{align}
where $\Tilde{P_\eta}$ is the  momentum of $\eta$ in the $\eta\eta$ rest frame,
\begin{align}
\Tilde{P_\eta}=\frac{\lambda^{1/2}(M^2_\mathrm{{inv}}(\eta\eta),m^2_\eta,m^2_\eta)}{2M_\mathrm{{inv}}(\eta\eta)}.
\end{align}

\section{Results}
\begin{figure}[H]
    \centering
\includegraphics[width=0.8\textwidth]{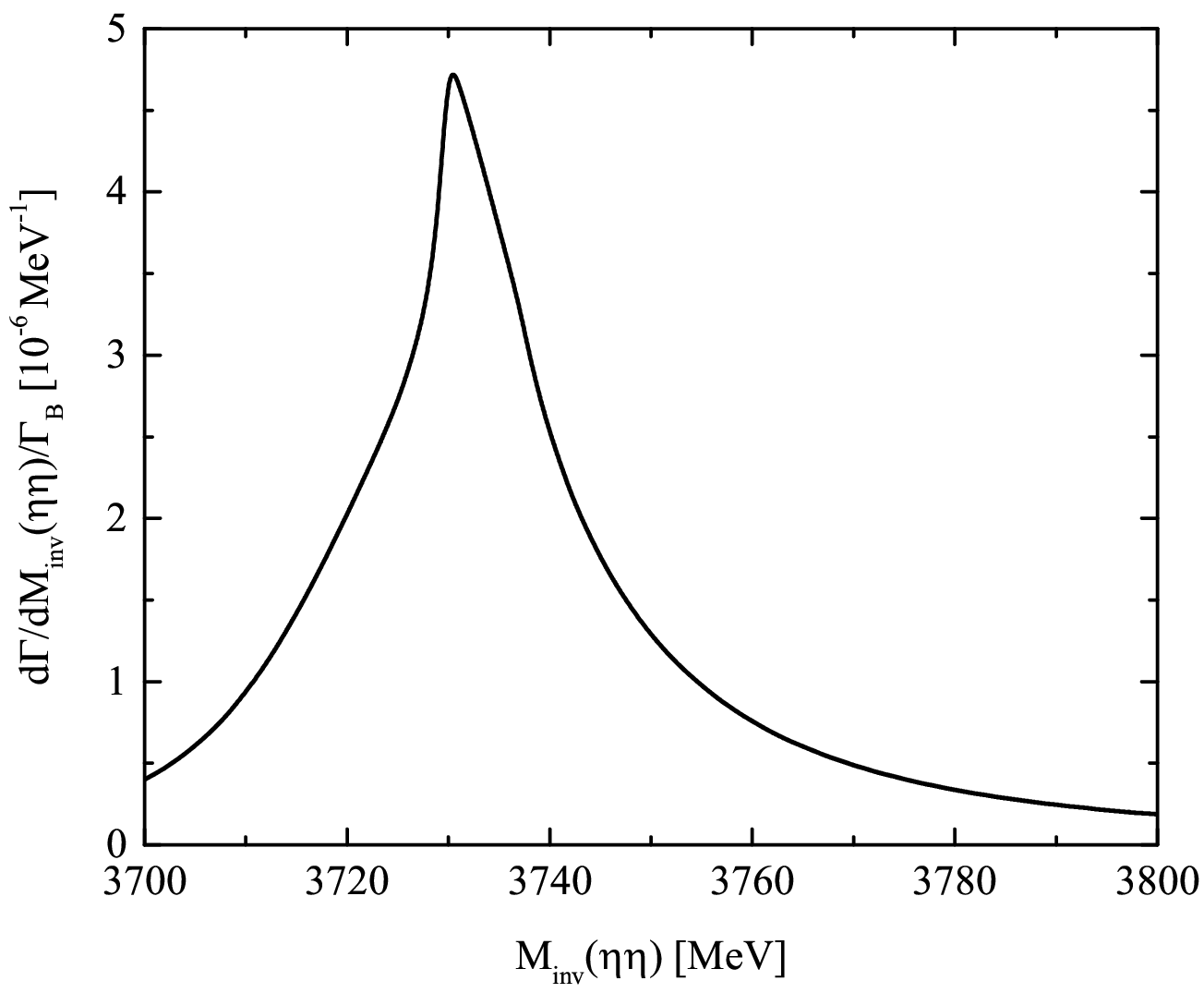}
    \caption{Results of $R_T$ as a function of $M_\mathrm{{inv}}(\eta\eta)$ invariant mass distribution.}
    \label{Minv}
\end{figure}

In Fig.~\ref{Minv} we show the results of $R_T=\frac{1}{\Gamma_B}\frac{d\Gamma}{dM_\mathrm{{inv}}}$. We make use of the value of the ratio $\frac{C^2}{\Gamma_B}$  from Eq.~(\ref{Csover}), hence we can   predict not only   the shape of the mass distribution but also its strength.   We indeed find in Fig.~\ref{Minv} a neat peak around the mass of the $D\Bar{D}$ bound state with the width predicted in Ref.~\cite{Xiao:2012iq}.
In order to find the feasibility of this experiment we integrate $R_T$ over $M_\mathrm{{inv}}(\eta\eta)$ to get the strength of the peak as $\frac{1}{\Gamma_B}\int \frac{d\Gamma}{dM_\mathrm{{inv}}(\eta\eta)}dM_\mathrm{{inv}}(\eta\eta)$, and we obtain the following value for the branching ratio of the reaction $B^+\to K^+D\Bar{D}|_b$; $D\Bar{D}|_b\to\eta\eta$, where $D\Bar{D}|_b$ means the $D\Bar{D}$ bound state,
$$Br[B^+\to K^+D\Bar{D}|_b; ~D\Bar{D}|_b\to\eta\eta]=1.47\times 10^{-4}$$
Taking into account that most of the hadronic branching fractions reported in the PDG are of the order of $10^{-4}$ or smaller, with some branching ratios of the order of $10^{-7}$, this branching ratio is relatively big and could easily be observed in experiments. This result can only encourage experimental teams to perform this measurement that would show for the first time the peak associated to the $D\Bar{D}$ bound state.

\section{Conclusions}

  We have studied the reaction $B^+\to K^+\eta\eta$ with the aim of finding a peak in the $\eta\eta$ mass distribution corresponding to a $D \Bar{D}$ bound state that has been predicted by several theoretical frameworks, in lattice QCD simulations, and has also been claimed to exist from  the observation of a concentration of strength around the $D \Bar{D}$ threshold in reactions producing $D \Bar{D}$ in the final state.\\
  
     In order to maximize the chances of observation we have selected a reaction that in a first step produces a $D \Bar{D}$ which is allowed to interact and produce the $\eta\eta$ at the end. The reaction chosen is $B^+\to D_s^{*+} \Bar{D}^0$, which has a large branching fraction for a $B$ hadronic decay, of the order of $10^{-2}$. The $D_s^{*+}$ decays to $D^0 K^+$ and the $D^0 \Bar{D}^0$ interact and produce the $\eta\eta$. Technically the combined process is evaluated  by means of a triangle diagram where the $D \Bar{D}$ are virtual, a necessary condition to produce the $D \Bar{D}$ bound state. The choice of $\eta\eta$ being produced by the $D^0 \Bar{D}^0$ interaction is motivated because the $D \Bar{D}$ bound state only decays in light meson pairs, where the $c \Bar{c}$ quarks have been annihilated. From previous calculations one knows that the $\eta\eta$ channel is one of the light pseudoscalar channels that couples most strongly to the $D \Bar{D}$ bound state. \\

     With this promising scenario we have evaluated the $\eta\eta$ mass distribution for the $B^+ \to K^+\eta\eta$ decays and we have found indeed a clear peak around the predicted mass of the $D \Bar{D}$ bound state.  Then we have integrated the mass distribution and found a branching fraction for $B^+ \to K^+$ ($D\Bar{D}$, bound); ($D\Bar{D}$, bound) $\to\eta\eta$ of the order of $1.5 \times 10^{-4}$. This is a relatively large branching fraction for a $B$ decay, which should encourage its search to finally find a peak for this much searched for state.\\
     
Final Note: While sending the paper to the Inspire web, a similar paper 
appeared there~\cite{Li:2023nsw} dealing on a similar reaction,
$B^-\to K^-\eta\eta_c$ decay. While the reaction is also promising, the 
method and formalism used in~\cite{Li:2023nsw} are different and no absolute 
rate is predicted.

\section{acknowledgements}

The work of P.C.S.B and L.M.A. is partly supported by the Brazilian agencies CNPq (Grant Numbers 309950/2020-1, 400215/2022-
5, 200567/2022-5), FAPESB (Grant Number INT0007/2016) and CNPq/FAPERJ under the Project INCT-F\'{\i}sica Nuclear e
Aplicações (Contract No. 464898/2014-5).
This work of J. S. is partly supported by the National Natural Science Foundation of China under Grants No. 12247108 and the China Postdoctoral Science Foundation under Grant No. 2022M720359. 
This work is
also partly supported by the Spanish Ministerio de Economia y Competitividad (MINECO) and European FEDER
funds under Contracts No. FIS2017-84038-C2-1-P B, PID2020-112777GB-I00, and by Generalitat Valenciana under
contract PROMETEO/2020/023. This project has received funding from the European Union Horizon 2020 research
and innovation programme under the program H2020-INFRAIA-2018-1, grant agreement No. 824093 of the STRONG-2020 project. This research is also supported by the Munich Institute for Astro-, Particle and BioPhysics (MIAPbP)
which is funded by the Deutsche Forschungsgemeinschaft (DFG, German Research Foundation) under Germany’s
Excellence Strategy-EXC-2094 -390783311.

\bibliography{refs.bib}
\end{document}